\documentclass[12pt]{aastex}

\slugcomment{}

\shorttitle{}
\shortauthors{Osten et al.} 

\begin{document}
\title{Connecting Flares and Transient Mass Loss Events in Magnetically Active Stars} 
\shorttitle{Connecting Stellar Flares and CMEs} 
 
\author{Rachel A. Osten\altaffilmark{1}}
\affil{Space Telescope Science Institute} 
\affil{3700 San Martin Drive, Baltimore, MD 21218}
\altaffiltext{1}{Also at Center for Astrophysical Sciences, Johns Hopkins University, Baltimore, MD 21218}
\email{osten@stsci.edu}

\and
\author{Scott J. Wolk}
\affil{Harvard-Smithsonian Center for Astrophysics}
\affil{60 Garden Street, Cambridge MA 02138}


\begin{abstract}
We explore the ramification of associating the energetics of extreme magnetic reconnection
events
with transient mass loss in a stellar analogy with solar eruptive events.
We establish energy partitions relative to the total bolometric radiated flare energy
for different observed components of stellar
flares, and show that there is rough agreement for these values with solar flares. 
We apply an equipartition between the bolometric radiated flare energy
and kinetic energy in an accompanying mass ejection, seen in 
solar eruptive events and expected from reconnection.
This allows an integrated flare rate
in a 
particular waveband to be used to estimate the amount of associated transient mass loss.
This approach is supported by a
good correspondence between observational flare signatures
on high flaring rate stars and the Sun, which suggests a common physical origin.
If the frequent and extreme flares that young solar-like stars and
low-mass stars experience are accompanied by transient mass loss in the form of 
coronal mass ejections, then the cumulative effect of this mass loss could be large.
We find that for young solar-like stars and
active M dwarfs, the total mass lost due to transient magnetic eruptions 
could have significant impacts on
disk evolution, and thus planet formation, and also exoplanet habitability. 
\end{abstract}


\section{Introduction}
Changes in the magnetic configuration of the Sun's outer atmosphere occur on short timescales
and lead to the release
of free energy which powers particle acceleration, plasma heating, shocks and mass motions.
These solar eruptive events manifest themselves as flares, coronal mass ejections (CMEs), and proton events.
Flares have been observed on many different kinds of cool stars, ranging from 
F dwarfs \citep{mm2000},
G and K giants \citep{ayres1999,ayres2001,testa2007},
tidally locked binary systems \citep{osten2004}, dMe flare stars \citep{hawley1995},
hyperactive young Suns containing star-disk interactions \citep{favata2005},
and very low mass stars near the substellar limit \citep{stelzer2006}.  
Magnetic reconnection processes are thought to underly the events, both on cool stars
and the Sun 
\citep{shibata1999,shibatayokoyama1999}.
Detailed studies of various aspects of stellar flare phenomena show global agreement with processes
occurring during solar flares, implying a similar physical origin, despite the disparity
in scale -- the largest stellar flares can be up to one million times more energetic than the largest
solar flares \citep{osten2007}.
Stellar counterparts to solar flare phenomena include white light stellar flares \citep{hawley1995}, soft X-ray
flares \citep{reale1997}, transition region and chromospheric flares \citep{hawley2003},
nonthermal gyrosynchrotron emission \citep{osten2005},
nonthermal hard X-ray emission \citep{osten2007}, and coherent radio emission \citep{ostenbastian2006}.
This diversity underscores the multi-wavelength nature of the flare process; 
the 
details connecting the energetics 
in the different wavelength regions is still unclear.

While flares on cool stars are observed across the electromagnetic spectrum, the
detection of transient mass loss signatures in cool dwarf stars (as with steady mass loss signatures)
remains a difficult prospect \citep{leitzinger2011}.
\citet{houdebine1996} and references therein list a number of outflows detected in stellar
flares, but there are more redshifted than blue-shifted emission components, and the hetergeneous
nature of the events prevents a systematic exploration. Thus, in the absence of
conclusive evidence of stellar CMEs, stellar astronomers must look to the Sun
to gain insight into transient stellar mass loss on magnetically active stars.

The commonality of stellar flare phenomenology with solar observations suggests that an extrapolation of
solar flare/CME behavior to the more distant stars should be appropriate.
The Sun has a low overall rate of mass loss ($\dot M_{\rm \odot}=2\times
10^{-14}$ M$_{\odot}$ yr$^{-1}$), of which transient mass loss events comprise a minor
component \citep[typically $<$10\%;][]{howard1985}.
However, if mass loss due to CMEs scales with flare occurrence, 
stars with a high flaring rate ought to also show an enhanced rate of mass lost due to
transient events like CMEs.
Recently, 
\citet{aarnio2012} used an empirical scaling between solar flare X-ray energy and associated CME mass
and extrapolated to pre-main sequence stars with measured X-ray flare energies to deduce mass loss rates of
putative CMEs associated with the stellar flares. They inferred high levels of CME-related mass loss
in these solar-type pre-main sequence stars of 10$^{-12}$-10$^{-9}$ M$_{\odot}$ yr$^{-1}$.
\citet{drake2013} also used an empirical relationship between solar flare X-ray energy and associated
CME mass to extrapolate to observed stellar coronal X-ray luminosities and investigate the 
radiative and kinetic energy requirements, assuming that the stellar
coronal emissions are produced by flares whose occurrence has a power-law dependence on flare energy.
They likewise find a large CME mass loss rate.
The two studies were limited to the X-ray spectral region due to the large databases of solar flares
measured at X-ray wavelengths, i.e. the GOES 1-8 \AA\ band, and the fairly large databases of 
stellar X-ray flares which have accumulated over the few decades of sensitive X-ray astronomical
telescopes.  
However, solar and stellar flare studies do take place in other wavelength regions, notably
optical (see \citet{hawleypettersen1991} for studies of M dwarf flares and the recent results of \citet{maehara2012} for studies of energetic flares on solar-like stars).
Flare frequency distributions have long been constructed for optical wavelength flare emissions
\citep{lme1976},
and a different method must be employed to investigate possible transient mass loss associated with 
stellar flares studied in these wavelength regions.
In addition, recent solar flare studies have demonstrated that the X-ray band is a minor contributor to the total radiated
flare energy integrated over all wavelengths, the bolometric flare energy \citep{woods2006,kretzschmar2011}.
Multiwavelength stellar flare studies have also shown that coronal energy flare losses are dwarfed by 
optical flare emissions \citep{hawley1995}.
Taking these factors into account requires
a better way to intercompare solar and stellar flare energies for a more
robust determination of the flare bolometric energy. 
Extrapolating solar flare energy and transient mass loss to the stellar regime also
needs a way to connect
flares and CMEs that goes beyond empirical correlations, to avoid the pitfall of ``big'' scaling with ``big'' without a basis for the relation.
In this paper we explore stars with high flaring rates using studies
from disparate wavelength regions and probe the implications for 
enhanced stellar mass-loss rates using a physically motivated relationship between flares and CMEs.


\section{Flare Energy Partition }

Although there have been many more multi-wavelength studies of solar flares than their
stellar counterparts,
a synthesis of recent results demonstrates similar patterns of energy partition for the coronal plasma and 
optical continuum emission in both solar and stellar flares. Table~\ref{tbl:ex} explains a number of
energy contributions discussed in the paper.

The existence of solar white-light flares has been known for years, yet it was only with the study of \citet{woods2006}
that a solar flare was detected in an equivalent manner to stellar observations, e.g. 
in integrated solar light (Total Solar Irradiance or TSI). 
In that study the peak intensity of the flare in TSI was only 268 ppm above the non-flare
emission, compared with
a factor of 570 increase in the Geostationary Operational Environment Satellite (GOES) 1-8 \AA\ band. 
The flare in the X-ray band, while exhibiting a much larger peak flux enhancement,
contributed only $\sim$1\% to the bolometric radiated flare energy: 
E$_{\rm bol}$/E$_{\rm X}\approx$105.
The study of \citet{kretzschmar2011} 
extended this result to fainter solar flares by shifting and
co-adding the TSI signal corresponding to the times of GOES flares of different peak intensities. 
They found a range of values for  $TSI/GOES$ energy ratios,
from 90$\pm$10 for the strongest flares, to 330$\pm$130 for the flares that are up
to 400 times weaker. 
The coronal flare radiation occurs over a wider wavelength range than the GOES band, however, and
integrating the radiative contribution of coronal plasma  
over a wider band to encompass most of this reveals that the total solar coronal flare radiated energy
is 20\% of the bolometric flare energy \citep{emslie2012}, $E_{\rm cor}/E_{\rm bol}=0.2$.
\citet{kretzschmar2011}
also demonstrated that the white light solar flare emission from these disparate
flare classes
has a continuum shape consistent with that of a black-body spectrum
at about 9000 K,  which contributes $\sim$70\% of the total radiated energy.
These solar flare studies reveal that the two main contributors to the bolometric flare energy
are the optical continuum emission and soft X-ray radiation.

Multi-wavelength studies of stellar flares are fewer in number, yet
reveal basic similarities as to the energy partition.
The radiative energy budget for flares on M dwarfs in the UV-optical wavelength range has been established
by \citet{hawleypettersen1991}: 
the continuum component 
has a shape which is characterized by a blackbody with a temperature
at peak of 9000$-$10,000 K. 
This component dominates over line radiation
when summed over all phases of the flare, carrying 90\% of the optical radiated energy; 
$E_{\rm cont}/E_{\rm opt}$=0.9.
Due to the high temperature of this continuum component, the U-band is commonly used for photometric monitoring and is found 
to contain about 16\% 
of the total optical energy
\citep{hawleypettersen1991},
$E_{\rm U}/E_{\rm opt}$=0.16.

\citet{hawleypettersen1991} established the dominance of the blackbody continuum 
in the optical/ultraviolet part of the stellar flare energy budget.  Further observations of
the optical and X-ray components of a stellar flare measured simultaneously (and with complete
temporal coverage of the flare in both bands) establish the importance of the coronal flare contribution
to the bolometric flare radiated energy, assuming that these two components are the 
major contributors to the flare radiated energy.
\citet{hawley1995} studied an event on the nearby flare star AD~Leo
(``EF2'') which had optical and X-ray coverage.
The energy in the U band filter integrated over all phases of the flare is 2.6$\times$10$^{32}$
erg (their Table 2), and using the result above for the fraction of total optical flare energy that appears in the 
U band, gives a total optical flare energy of 1.6$\times$10$^{33}$ erg. 
The integrated flare energy in the EUV bandpass of 65-190 \AA\ was 8.2$\times$10$^{31}$ erg (their Table 2).
In order to correct this for the total coronal flare radiated over a broader bandpass and determine $E_{\rm cor}$, 
we use flare temperature and
emission measure values determined from this flare 
(peak temperature of 2$\times$10$^{7}$ K, emission measure of 2.5$\times$10$^{51}$
cm$^{-3}$) and determine the fraction of E$_{\rm cor}$
occurring in the 65-190 \AA (0.065-0.19 keV) band
relative to a wider band of 0.01-10 keV to be 0.12, and for the GOES bandpass of 1-8 \AA\ relative to this wider energy band, to be
0.29. 
This leads to a total coronal radiated flare energy $E_{\rm cor}$ estimate of 6.8$\times$10$^{32}$ erg. 
This is to be
 compared against the total optical flare energy of 1.6$\times$10$^{33}$ erg, for a total bolometric
flare radiated energy $E_{\rm bol}=E_{\rm cor} + E_{\rm opt}$ of 2.28$\times$10$^{33}$ and a fraction 
$E_{\rm cor}/E_{\rm bol}$=0.3.
The two dominant components to the bolometric flare energy are the optical/UV emission 
arising from the photosphere/chromosphere and the high energy emission from the corona:
this implies $E_{\rm opt}$/$E_{\rm bol}=$0.7. 
Then the U band carries a fraction $E_{\rm U}$/$E_{\rm bol}$=0.11. 
Since the energy in the hot optical continuum component $E_{\rm cont}$
comprises 90\% of the optical energy, this 
also suggests that $E_{\rm cont}/E_{\rm bol}=$0.6, similar to the solar flare ratio of 0.7.


Based on the good agreement between these energy partition values
for solar flares and the stellar flare 
discussed above, we assume that these numbers are applicable to all flares
to estimate the bolometric radiated
flare energy for flares occurring on stars of differing types. 
There is a paucity of multi-wavelength flare campaigns
on stars of differing types, and it is possible that the ratios could vary from flare to flare (or star to star), but these
numbers are a starting point for such investigations.
We note that \citet{hawley2003} found that the optical and ultraviolet emission properties and energy budgets of a 
sample of different types of flares were remarkably similar, which supports this approach.
Table~\ref{tbl:f} lists these $f=E_{\rm rad}/E_{\rm bol}$ conversion factors,
which estimate the fraction of energy released in the band considered relative to the total bolometric radiated
flare energy.
For stellar flares observed with the Kepler mission, 
and using the tabulated Kepler bandpass response in the Kepler Instrument Handbook \citep{kihb}, we compute the fraction of the energy from
a hot blackbody continuum which appears in this wavelength range, using a blackbody temperature
of 9000 K and a range of $4000-9000$ \AA; the
fraction $E_{\rm cont, Kepler}/E_{\rm cont}$=0.26.
These energy calculations for solar and stellar cases do not take into account the contribution to the total
flare energetics of other phenomena known to occur along with optical and coronal radiation, 
particularly the kinetic energy contained in accelerated particles, so they do not represent a true flare
energy budget, but they are used to estimate a bolometric radiated flare energy to place disparate
solar and stellar flare studies on a more common footing.


\begin{table}[h]
\caption{Different Energy Contributions Discussed in the Paper \label{tbl:ex}}
\begin{tabular}{ll}
\hline
E$_{\rm rad}$ & flare radiated energy in a particular wavelength region \\
E$_{\rm bol}$ & flare radiated energy integrated over all wavelengths\\
E$_{\rm cont}$ & flare radiated energy in the blackbody continuum \\
E$_{\rm cont, Kepler}$ & flare radiated energy in the blackbody continuum which occurs in the Kepler bandpass\\
E$_{\rm opt}$ & flare radiated energy in the optical and UV, includes line and continuum radiation\\
E$_{\rm U}$ & flare radiated energy in the U filter bandpass\\
E$_{\rm cor}$ & flare radiated energy from coronal plasma \\
E$_{\rm GOES}$ & flare radiated energy from coronal plasma in the GOES bandpass\\
E$_{\rm CME}$ & total energy in the coronal mass ejection\\
E$_{\rm KE, CME}$ & kinetic energy in the coronal mass ejection\\
\hline
\end{tabular}
\end{table}

\begin{table}[h]
\caption{Conversion factors for radiated flare energies in different bandpasses relative to
the bolometric flare radiated energy \label{tbl:f}}
\begin{tabular}{llll}
\hline
bandpass & wavelength range & value of $f=E_{\rm rad}/E_{\rm bol}$ & value of $f=E_{\rm rad}/E_{\rm bol}$ \\
	 &                  &  (Sun)  				   & (active stars) \\
\hline
GOES & 1-8 \AA\ & 0.01 & 0.06\\
SXR &0.01-10 keV &0.2 &  0.3 \\ 
hot blackbody &1400-10000 \AA\  &0.7 &  0.6 \\
U &3000-4300 \AA\ &  &0.11 \\
Kepler & 4000-9000 \AA\ & & 0.16$^{\dagger}$\\
\hline
\multicolumn{4}{l}{$^{\dagger}$ assuming T$_{\rm BB}=$9000 K}
\end{tabular}
\end{table}

\section{Connecting Flares and Coronal Mass Ejections }

\subsection{Equipartition between CME Energy and Flare Radiated Energy}
A rough equipartition between flows and radiation is a natural consequence if magnetic reconnection is the driver of the eruptive event
\citep{priestforbes}.
Recent solar studies corroborate this expectation.
\citet{emslie2004} investigated the detailed energy partition in two well-studied flare/CME
pairs which had constraints on thermal coronal energy, energy in nonthermal electrons and ions, and CME kinetic and
potential energy, and energetic interplanetary particles.
They concluded that the CME kinetic energy dominated over the other energies quantified, by large
factors.  Based on the later determination that total solar radiated flare energies can exceed those measured
in the X-ray by large factors,
\citet{emslie2005} used refinements to the amount of radiated flare energy in \citet{emslie2004} to show
that although the CME contains the greatest fraction of released energy, the uncertainties in the
different flare components and increased estimates of total radiative energies put the mechanical
energy of the CME at a rough equipartition with the total radiated energy in the flare. 
A more recent study of the global energetics of a larger sample of solar eruptive events
\citep{emslie2012} confirms this relationship; the bolometric radiated flare energy, or
the flare energy radiated over all wavelengths from all components of the flare, was 
about one third of the total energy
of the mass ejection for the 38 events studied.
The kinetic energy of the CME dominates the 
mechanical energy, $E_{\rm CME}$ $\approx$ E$_{KE,CME}$.

For large solar flares, there is a high rate of association between the occurrence of a flare
and an accompanying coronal mass ejection \citep{yashiro2006}.
Because the solar flare model appears to be validated in active stars, we expect that the
condition of 
rough equipartition between the energy in the CME and the bolometric flare radiated energy
in solar eruptive events should also hold in active stars, and use this physical model to
explore the cumulative effect of transient mass loss in high flaring rate stars.
Then, if E$_{\rm CME} \approx E_{\rm KE,CME}=1/2 Mv^{2}$ and
$\epsilon=E_{\rm bol}/E_{\rm CME}$, we constrain the mass of the ejection from the
energy in the flare, given an assumption about the velocity of the ejected material. 
The velocity of stellar CME is unknown, but is taken to be the escape velocity 
of the star. 
The energy partition established in the previous section 
is used to correct the observed radiated flare energy $E_{\rm rad}$ to 
its bolometric value by  $E_{\rm bol}= E_{\rm rad}/f$.

Flare energy distributions are commonly made for both the Sun and other stars; they are 
generally described by a power-law distribution of the form
\begin{equation}
\label{eqn:dnde}
\frac{d \dot N}{dE_{\rm rad}}=K_{E} \;\; E_{\rm rad}^{-\alpha}
\end{equation}
where $d\dot N/dE_{\rm rad}$ gives the differential number of flares occurring per unit time per unit energy,
and E$_{\rm rad}$ refers to the measurement of flare energies in a particular band of the
electromagnetic spectrum. The power-law is generally considered valid over the extent $E_{\rm rad,min}$
to E$_{\rm rad, max}$.
Measured values of $\alpha$ in stars and the Sun range from about 1.7 to 2.3 \citep{gudel2007}.

By expressing the number of CMEs per unit time per unit mass range as $d \dot N/dM$, the
total mass lost per unit time due to CMEs is determined using
\begin{equation}
\label{eqn:dndm}
\dot M_{\rm CME}=\int_{M_{\rm CME,min}}^{M_{\rm CME,max}}  M_{\rm CME} \left( \frac{d \dot N}{dM} \right) dM \;\;\; .
\end{equation}
with the limits of integration corresponding to the minimum and maximum CME mass.
Assuming that the equipartition between CME kinetic energy and total radiated flare energy
applies to most flares and associated CMEs, \\
\begin{eqnarray}
E_{\rm KE, CME} = \frac{E_{\rm bol}}{\epsilon} \\
M_{\rm CME}v^{2}/2  =  \frac{E_{\rm rad}}{\epsilon f} \;\;\; ,
\end{eqnarray}
or M$_{\rm CME} = 2 E_{\rm rad}/(\epsilon f v^{2})$, the total mass lost
in CMEs per unit time is, using  equations 2, 3 and 4, \\
\begin{equation}
\dot M_{\rm CME}=\int_{E_{\rm rad,min}}^{E_{\rm rad,max}} \frac{2 E_{\rm rad}}{v^{2}f\epsilon} \frac{d \dot N}{dE_{\rm rad}} \; dE_{\rm rad} \; .
\end{equation}
We have assumed a constant linear speed $v$ for all CMEs and we have related the limits of integration of mass
in the first equation to the corresponding minimum and maximum flare energy of the associated flare.
Inserting the relationship from equation~\ref{eqn:dnde} for the flare frequency distribution with energy leads to \\
\begin{equation}
\dot M_{\rm CME}=\int_{E_{\rm rad,min}}^{E_{\rm rad,max}} \frac{2 E_{\rm rad}}{v^{2}f\epsilon} K_{E} E_{\rm rad}^{-\alpha} \; dE_{\rm rad}.
\label{eqn:eqn6}
\end{equation}
It is intuitively clear that the cumulative transient mass loss rate should be related to
the total flare rate. 
This, it turns out, will allow us to simplify equation~\ref{eqn:eqn6}. 
The total flare rate per unit time can be expressed as \\
\begin{eqnarray}
\dot N_{\rm tot} = \int_{E_{\rm rad,min}}^{E_{\rm rad,max}} \frac{d \dot N}{dE_{\rm rad}} \;\; dE_{\rm rad}\\
 = \int_{E_{\rm rad,min}}^{E_{\rm rad,max}} K_{E} \;\; E_{\rm rad}^{-\alpha} \; dE_{\rm rad} \;\;, \\
 = \frac{K_{E}}{1-\alpha} \left(   E_{\rm rad,max}^{1-\alpha} - E_{\rm rad,min}^{1-\alpha }\right)
\end{eqnarray}
Evaluating the integral in equation 6 from a minimum flare energy $E_{\rm rad,min}$ to
maximum flare energy $E_{\rm rad,max}$, and using the result of equation 9 for the normalization,
the estimated total mass lost due to CMEs per unit time is then
\begin{equation}
\label{eqn:mdot}
\dot M_{\rm CME}=\frac{2}{v^2} \frac{\dot N_{\rm tot}}{\epsilon f} \frac{(1-\alpha)}{(2-\alpha)}
\frac{(E_{\rm rad,max}^{2-\alpha}-E_{\rm rad,min}^{2-\alpha})}{(E_{\rm rad,max}^{1-\alpha}-E_{\rm rad,min}^{1-\alpha})}
\end{equation}
where $\dot N_{\rm tot}$ is the number of flares per unit time occurring between $E_{\rm rad,min}$ and
$E_{\rm rad,max}$, and $E_{\rm rad}$ is the radiated energy in the particular wavelength band being
considered.
We recast equation~\ref{eqn:mdot} in a format which emphasizes the important variables; 
setting $R=E_{\rm rad,max}/E_{\rm rad,min}$ as the ratio of the maximum to minimum
flare energies observed, we see \\
\begin{equation}
\label{eqn:mdotsimple}
\dot M_{\rm CME} = \frac{2}{v^{2}} \frac{\dot N_{\rm tot}}{\epsilon f} E_{\rm rad,min} G(\alpha,R)
\end{equation}
where \\
\begin{equation}
G(\alpha,R)=\frac{1-\alpha}{2-\alpha} \left( \frac{R^{2-\alpha}-1}{R^{1-\alpha}-1} \right) \;.
\end{equation}
The value of $\alpha$ for a flare frequency distribution determine how  much weight is put on 
the largest and smallest flares; large values of $\alpha$ give more weight to
$E_{\rm rad,min}$ while smaller values of $\alpha$ do the opposite.  Because observed
values of $\alpha$ lie in a small range, 
G varies more due to R than with $\alpha$.
Table ~\ref{tbl:calc} lists values of 
 $\dot N_{\rm tot}$, $\alpha$, $E_{\rm rad, max}$ and $E_{\rm rad,min}$ 
from some flare frequency distributions, which together with flare conversion factors in Table~\ref{tbl:f} can be
used to calculate flare-associated transient mass loss.
\begin{deluxetable}{lccccccccc}
\rotate
\tabletypesize{\scriptsize}
\tablewidth{0pt}
\tablecolumns{10}
\tablecaption{Calculation of Transient Mass Loss Using Published Flare Energy Distributions \label{tbl:calc}}
\tablehead{ 
\colhead{Object} & \colhead{v$_{\rm esc}$} & \colhead{$\dot N_{\rm tot}$} & \colhead{E$_{\rm rad,min}$} &\colhead{E$_{\rm rad,max}$} & \colhead{band} &   \colhead{$\alpha$} 
& \colhead{$\dot M_{\rm CME}$}&  \colhead{$\dot M_{\rm CME,Aarnio}$} & \colhead{$\dot M_{\rm CME,Drake}$}\\ 
\colhead{} &  \colhead{km s$^{-1}$} & \colhead{flares star$^{-1}$ yr$^{-1}$}   & \colhead{erg} & \colhead{erg}  &  \colhead{}
& \colhead{}
 & \colhead{(M$_{\odot}$/yr)}   & \colhead{(M$_{\odot}$/yr)} & \colhead{(M$_{\odot}$/yr)}
}
\startdata
\multicolumn{8}{c}{-- Solar Mass Stars --}\\
Sun\tablenotemark{1}& 620  &200 & 5.6$\times$10$^{27}$ & 5.6$\times$10$^{30}$ & GOES &  1.79$\pm$0.05 &  4$\times$10$^{-16}$ & 8$\times$10$^{-16}$
&6$\times$10$^{-16}$ \\
Sun\tablenotemark{2}&  620  &  1577 & 5.6$\times$10$^{27}$ & 1.1$\times$10$^{30}$& GOES  & 2.03$\pm$0.09  & 1$\times$10$^{-15}$ &5$\times$10$^{-15}$ & 4$\times$10$^{-15}$\\
young suns in Orion\tablenotemark{3}& 440  & 51 & 2$\times$10$^{34}$ & 9$\times$10$^{36}$ & SXR & 1.66 &  2$\times$10$^{-11}$ & 4$\times$10$^{-13}$
& 2$\times$10$^{-13}$\\
EK Dra\tablenotemark{4}& 670  & 1550 & 3$\times$10$^{33}$& 3$\times$10$^{34}$ & SXR  & 2.08$\pm$0.34 & 9$\times$10$^{-12}$ & 2$\times$10$^{-12}$& 9$\times$10$^{-13}$\\
superflaring sun-like stars\tablenotemark{5} & 620 &3.5 & 5$\times$10$^{34}$ & 2$\times$10$^{36}$ & Kepler   & 2.3$\pm$0.3 & 8$\times$10$^{-13}$ &
3$\times$10$^{-14}$ & 2$\times$10$^{-14}$\\
\multicolumn{8}{c}{-- Low Mass Stars -- } \\
young low mass stars\tablenotemark{8}&214  &  9 & 2$\times$10$^{34}$ & 3$\times$10$^{35}$ & SXR & 2.2$\pm$0.2  &  
3$\times$10$^{-12}$ & 3$\times$10$^{-14}$ & 
2$\times$10$^{-14}$\\
AD Leo\tablenotemark{6}& 630 & 2557 & 6$\times$10$^{31}$ & 2$\times$10$^{33}$ & SXR & 2.02$\pm$0.28 &   5$\times$10$^{-13}$ & 3$\times$10$^{-13}$ & 2$\times$10$^{-13}$\\
AD Leo\tablenotemark{7} &630 & 3790 & 2$\times$10$^{30}$ & 3$\times$10$^{31}$ & U &  1.82$\pm$0.27 & 6$\times$10$^{-14}$ & 8$\times$10$^{-14}$ &
6$\times$10$^{-14}$\\
EV Lac\tablenotemark{6}& 610 & 1461 & 8$\times$10$^{31}$ & 2$\times$10$^{33}$ & SXR &  1.76$\pm$0.33 &  4$\times$10$^{-13}$ & 2$\times$10$^{-13}$ 
& 1$\times$10$^{-13}$\\
EV Lac\tablenotemark{7} & 610 &1790. & 1$\times$10$^{31}$&5$\times$10$^{31}$ & U &  1.69$\pm$0.11 &  9$\times$10$^{-14}$ & 9$\times$10$^{-14}$ &
6$\times$10$^{-14}$\\
inactive early M dwarfs\tablenotemark{9} & 630 & 26 & 8$\times$10$^{29}$ & 10$^{31}$ & U & 1.97$\pm$0.3 & 10$^{-16}$ & 3$\times$10$^{-16}$ &
2$\times$10$^{-16}$\\
inactive mid-M dwarfs\tablenotemark{9} & 630 & 50 & 4$\times$10$^{29}$ & 4$\times$10$^{32}$ & U &1.20$\pm$0.1 & 4$\times$10$^{-15}$ & 2$\times$10$^{-15}$ & 2$\times$10$^{-15}$\\
\enddata
\tablenotetext{1}{Taken from \citet{yashiro2006}.}
\tablenotetext{2}{Taken from \citet{veronig2002}.}
\tablenotetext{3}{Taken from \citet{wolk2005}.}
\tablenotetext{4}{Taken from \citet{audard1999}.}
\tablenotetext{5}{Taken from \citet{maehara2012}.}
\tablenotetext{6}{Taken from \citet{audard2000}.}
\tablenotetext{7}{Taken from \citet{lme1976}.}
\tablenotetext{8}{Taken from \citet{caramazza2007}.}
\tablenotetext{9}{Taken from \citet{hilton2011}.}
\end{deluxetable}

\subsection{Empirical Relationship between CME Mass and Flare X-ray Energy}

Both \citet{aarnio2012} and \citet{drake2013} derived empirical relationships between solar CME mass
and solar flare GOES energies, with a relationship of the form \\
\begin{equation}
\label{eqn:km}
M_{\rm CME} = K_{M} E_{\rm GOES}^{\gamma} \;,
\end{equation}
where $\gamma$ is a parameter with an empirical value of $\sim$0.6 in both studies.
Although it is not explicitly stated in \citet{aarnio2012}, $E_{G}$ in both cases
appears to be the flare energy in the GOES bandpass:
the figures in both \citet{aarnio2011} and \citet{aarnio2012}  use GOES
bandpass flare energies.  
We use these relationships to estimate a cumulative mass loss rate if the same empirical
relationship holds to the classes of flaring stars considered below, separated from solar flare energies
by orders of magnitude.
Because these stellar flare
compilations are not generally made in the GOES bandpass, we need to apply the flare
conversion factors appropriate for the bandpass in which stellar flares are compiled. If we denote
this radiated energy bandpass $E_{\rm rad}$ with a conversion factor $f_{\rm rad}$ to relate
to the bolometric flare energy, then equation~\ref{eqn:km} can be re-written \\
\begin{equation}
M_{\rm CME} = K_{M}^{'} E_{\rm rad}^{\gamma} \;\;\; ,
\end{equation} 
where \\
\begin{equation}
K_{M}^{'}=K_{M} \left( \frac{f_{\rm GOES}}{f_{\rm rad}} \right)^{\gamma} \;\;.
\end{equation}
Then the cumulative mass loss rate due to CMEs can be expressed according to equation~\ref{eqn:dndm},
and converting the relationship $d\dot N/dM$ into $d\dot N/dE_{\rm rad}$ using the relationship between
CME mass and flare energy, \\
\begin{equation}
 \dot M_{\rm CME} = \frac{\dot N_{\rm tot} (1-\alpha) K_{M}^{'}}{(E_{\rm rad,max}^{1-\alpha}-E_{\rm rad,min}^{1-\alpha})}    \int_{E_{\rm rad,min}}^{E_{\rm rad,max}} E_{\rm rad}^{\gamma-\alpha} \;\; dE_{\rm rad}
\end{equation}
or,
\begin{equation}
\label{eqn:mdot2}
\dot M_{\rm CME} = \frac{\dot N_{\rm tot} (1-\alpha) K_{M}^{'}}{(E_{\rm rad,max}^{1-\alpha}-E_{\rm rad,min}^{1-\alpha})}
\frac{1}{\gamma-\alpha+1} (E_{\rm rad,max}^{\gamma-\alpha+1}-E_{\rm rad,min}^{\gamma-\alpha+1}) \;\;\; .
\end{equation}
Similar to equation~\ref{eqn:mdotsimple} with $R=E_{\rm rad,max}/E_{\rm rad,min}$, we express this so that the dependencies are clearer:
\begin{equation}
\label{eqn:mdot2simple}
\dot M_{\rm CME} = \dot N_{\rm tot} K_{M}^{'} E_{\rm rad,min}^{\gamma} G_{2} (\gamma,\alpha,R)
\end{equation}
with 
\begin{equation}
G_{2}(\gamma,\alpha,R)= \frac{1-\alpha}{\gamma-\alpha+1} \left( \frac{R^{\gamma-\alpha+1}-1}{R^{1-\alpha}-1} \right)\;\;. 
\end{equation}

Using the parameters $\dot N_{\rm tot}$, $\alpha$, $E_{\rm rad, max}$ and $E_{\rm rad,min}$ listed in Table~\ref{tbl:calc}, and the conversion factors for the GOES band and the appropriate wavelength region in 
Table~\ref{tbl:f}, the mass lost due to CMEs using the relationships of
\citet{aarnio2011} and \citet{drake2013}  are also calculated. 
The values for $K_{M}$ and $\gamma$
from the \citet{aarnio2011} study are $K_{M}=(2.7\pm1.2)\times 10^{-3}$, $\gamma=0.63\pm0.04$,
and for \citet{drake2013} the parameters from their empirical fits are
$K_{M}=0.03 (0.01-0.1)$, $\gamma=0.59\pm0.02$.
The empirical equations \citeauthor{aarnio2011} and \citeauthor{drake2013} establish for solar CME masses
based on solar GOES flares return very similar values of mass for a given flare energy,
so their implied $\dot M_{\rm CME}$ should be very similar.
However the uncertainty on their fit parameters (particularly the constant $K_{M}$) 
is large, leading to a spread of $\approx$ 150-300 in the mass calculated from 
\citeauthor{drake2013}'s equation 1, and a spread of 600-2400 in the mass calculated from
\citeauthor{aarnio2012}'s equation 2.

\section{Evaluation and Application}

We have argued that we understand the distribution of flares as a function of energy. Combining this
with the assumption of a common energy partition between flare energy in a given bandpass to the total radiated energy,
and that a single constant speed is appropriate to be applied to all CMEs, we estimate the
amount of mass lost due to CMEs in a variety of flaring stars using a physically motivated
rationale for connecting flares and CMEs.
This method can also be applied to the Sun, where several calculations of  flare energy distributions
have been made, and where the total mass lost in CMEs can be determined through observations.
The flare energy partitions also allow the use of an empirical relationship between solar CME mass
and solar flare energy to be extrapolated over the several orders of magnitude difference between
solar and stellar flares and applied to stellar flares observed in a range of wavelength regions.

We take a constant value of CME speed equal to the
escape velocity of the star (for the Sun, this is $\approx$ 620 km s$^{-1}$, so not
that different from the average flare-associated speed in \citet{aarnio2011}).
We take $\epsilon=1$ to evaluate the case for an equipartition between flare bolometric radiated  energy
and CME kinetic energy, although we note that the $\epsilon=0.3$ determination of
\citet{emslie2012} would suggest a higher CME mass-loss rate by a factor of $\approx$3. 

Table~\ref{tbl:calc} lists parameters for flare frequency distributions from studies of several 
different types of stars, and the estimated total mass lost due to CMEs which might be associated with
them using the above two formulations (equations ~\ref{eqn:mdot} and ~\ref{eqn:mdot2}).

For the  flare energy distributions of stars discussed in the following subsections, we
also estimated the total mass lost using the same energy ranges for all stellar types considered.
This facilitates comparison of the technique using different stars and flares characterized over different
wavelength ranges.
\citet{yashiro2006} showed that above an X-ray (1--8 \AA) flare energy of 5$\times$10$^{29}$ erg, 
all solar flares in their study had CMEs associated with them. 
Below this level the flare-CME association on the Sun breaks down, for reasons which
are not yet understood.   This flare energy level corresponds to an $E_{\rm bol}$ of $\sim$5$\times$10$^{31}$ erg.
\citet{karpen2012} describe the onset of solar CMEs as being due to the start of
fast reconnection in a flare current sheet, which triggers explosive energy release
and ejection.  They comment that the available magnetic free energy 
determines whether a system ends up erupting, which, assuming that the final CME and flare energies
traces the inital magnetic energy, would agree with the decrease in 
total magnetic energy available in events with decreasing  
CME and flare energies. 
We take this lowest bolometric radiated energy at which solar flares and CMEs
start to exhibit a break in close association as indicative 
of the point where there might not be enough free energy in the system for the
mass to eject from the star.
As discussed in \citet{schrijver2012}, the maximum flare energies observed on active
stars appears to be $\approx$10$^{37}$ erg. 
We use an E$_{\rm bol,min}$ of 5$\times$10$^{31}$ erg and an E$_{\rm bol,max}$ of 10$^{37}$ erg, and
define $R_{\rm bol}=E_{\rm bol,max}/E_{\rm bol,min}$ and $R_{\rm FFD}=E_{\rm bol,max,FFD}/E_{\rm bol,min,FFD}$.
Then the new $\dot N_{\rm tot}$ between E$_{\rm bol,min}$ and E$_{\rm bol,max}$ is \\
\begin{equation}
\label{eqn:mdotextrap}
\dot N_{\rm tot} = \dot N_{\rm tot,FFD} \left( \frac{E_{\rm bol,min}}{E_{\rm bol,min,FFD}} \right)^{1-\alpha}  \frac{(R_{\rm bol}^{1-\alpha}-1)}{(R_{\rm FFD}^{1-\alpha}-1)}
\end{equation}
where $\dot N_{\rm tot,FFD}$ is the total number of flares star$^{-1}$ yr$^{-1}$ between $E_{\rm bol,minFFD}$ and $E_{\rm bol,maxFFD}$
from the flare frequency distributions (FFDs) listed in Table~\ref{tbl:calc}, 
after converting $E_{\rm rad,min}$ and $E_{\rm rad,max}$ to their bolometric
equivalents 
($E_{\rm bol,minFFD}=E_{\rm rad,min}/f$ and similarly for
$E_{\rm bol,maxFFD}$).
These numbers are listed in Table~\ref{tbl:extrap}.

\begin{deluxetable}{lccc}
\tablewidth{0pt}
\tablecolumns{4}
\tablecaption{Extrapolation of Transient Mass Loss to Common Energy Ranges\tablenotemark{1} \label{tbl:extrap}}
\tablehead{ 
 \colhead{Object} &  \colhead{band} &  \colhead{$\dot N_{\rm tot}$} &
\colhead{$\dot M_{\rm tot,CME}$} \\
\colhead{} & \colhead{} & \colhead{\# flares yr$^{-1}$ star$^{-1}$} & \colhead{M$_{\odot}$ yr$^{-1}$}
}
\startdata
\multicolumn{4}{c}{-- Solar Mass Stars -- } \\
young suns in Orion&SXR &6$\times$10$^{3}$ & 4$\times$10$^{-11}$\\
EK Dra& SXR &5$\times$10$^{5}$ & 4$\times$10$^{-11}$\\
superflaring sun-like stars &Kepler &3$\times$10$^{5}$ &10$^{-11}$ \\
\multicolumn{4}{c}{-- Low Mass Stars -- } \\
young low mass stars&SXR &5$\times$10$^{4}$ &2$\times$10$^{-11}$\\
AD Leo&SXR &10$^{4}$ & 10$^{-12}$\\
AD Leo & U& 2$\times$10$^{3}$&10$^{-12}$ \\
EV Lac& SXR &6$\times$10$^{3}$  & 7$\times$10$^{-12}$\\
EV Lac & U&4$\times$10$^{3}$ &10$^{-11}$ \\
inactive early-M dwarfs &U & 4 & 8$\times$10$^{-16}$ \\
inactive mid-M dwarfs &U & 40 & 5$\times$10$^{-11}$ \\
\enddata
\tablenotetext{1}{Using parameters ($f$, $\alpha$, v$_{\rm esc}$) noted
in Table~\ref{tbl:calc}, and $N_{\rm tot}$ from Table~\ref{tbl:calc} adjusted to the common energy range 
of E$_{\rm bol,min}=$5$\times$10$^{31}$ erg and E$_{\rm bol,max}=$10$^{37}$ erg.}
\end{deluxetable}



\subsection{Solar-like Stars}

\subsubsection{Sun}
For application to the Sun, we take the compilations of solar flare frequency distributions
of \citet{yashiro2006} for flares with 
associated CMEs, and the flare frequency distribution of \citet{veronig2002}; parameters are listed in Table~\ref{tbl:calc}.
In the \citet{yashiro2006} paper, for flare energies below those with a 100\% association with CMEs, the frequency distribution is
corrected by the observed association.
Both papers use the {\it GOES} satellites for full-disk 1-8 \AA\ solar X-ray variations, and so 
we take the ratio $f$ to be 0.01, based on the total
solar irradiance measurements of \citet{woods2006}, to correct these energies to a bolometric radiated flare energy.
The solar flare study of \citet{veronig2002}
compiled distributions for solar flares over a 24 year time interval, without
regard to CME association. This long time span also covers two solar activity cycles, important because
of the variation of flare and CME occurrence with phase of the solar cycle.
They found a power-law of $\alpha=$2.03$\pm$0.09 between 5.6$\times$10$^{27}$ and 
1.1$\times$10$^{30}$ erg. Their total number of flares fitted was 37,851 flares in 24 years, for
1577 flares yr$^{-1}$. These parameters are listed in Table~\ref{tbl:calc}.
\citet{yashiro2006} fit a power-law of $\alpha=$1.79$\pm$0.05 between 5.6$\times$10$^{27}$ and
5.6$\times$10$^{30}$ erg to solar flares with associated CMEs, with $\sim$10$^{6}$ flares/(J m$^{-2}$) above
their lower energy limit of 2$\times$10$^{-3}$ J m$^{-2}$, or 5.6$\times$10$^{27}$ erg.
The time span of their study was 10 years, so we estimate $N_{\rm tot}$ in Table~\ref{tbl:calc}
to be 200 flares yr$^{-1}$.

These two studies return similar estimates for the cumulative amount of mass lost due to CMEs. 
The two total transient mass loss rates, 4$\times$10$^{-16}$ and 10$^{-15}$ M$_{\odot}$ yr$^{-1}$, are 
2\% and 6\%, respectively,
of the overall $\dot M_{\odot}$.
\citet{howard1985} showed that the average CME mass flux at Earth during a 4 year period (1978--1981)
was 5\% of the total solar wind mass loss.
\citet{webb1987} demonstrated that the CME mass flux at Earth shows solar cycle variations, 
containing 10\% of the total mass loss rate near solar maximum, and 1\% near solar minimum,
while \citet{jacksonhoward1993} demonstrated that 16\% of the solar wind at the maximum of the Sun's activity cycle
comes from CME mass loss.
The numbers derived from the work of \citet{veronig2002} took into account all flares observed in the
stated energy range. 
 This overestimates the numbers of CMEs occuring with these flares, which suggests that
the overestimation should be only a factor of a few, and still within the previously observed
variation in the total solar mass loss rate occurring due to transient mass loss events.
This generally good agreement provides additional grounding for extrapolating these results to
higher stellar flare energies and suggests that if the same processes are occurring in solar
and stellar eruptive events, this overprediction of the transient mass loss rate might be biased 
high by only factors of a few.

\subsubsection{The Young Suns of Orion}
\citet{wolk2005} compiled a flare frequency distribution for young solar-like stars in the Orion Nebula Cluster
observed with the {\it Chandra X-ray Observatory}. 
They computed a distribution from the ensemble behavior 
of selected flares on stars with masses in the range 0.9-1.2 M$_{\odot}$ at an age of $\sim$ 1MY.
 Using the spectral parameters for a sample star
which had temperatures near the median values, we determined conversion factors for the spectral energy
distribution of the flaring plasma between 0.5--8 keV  (the range considered in the paper)
and 0.01-10 keV, which will encompass the majority
of the coronal radiated energy.  
The flare frequency distribution is fitted between $E_{\rm 0.5-8 keV}$=10$^{34}$
and 6$\times$10$^{36}$ ergs, which would correspond to a range 2$\times$10$^{34}$--9$\times$10$^{36}$ erg between 0.01-10 keV.
We use $f$ tabulated in Table~\ref{tbl:f} for this energy range to convert to the total bolometric flare energy.
Masses and radii for these objects are tabulated in \citet{wolk2005}, and we use the average $v_{\rm esc}=$440 km s$^{-1}$
in our calculations.
There were 27 flares above the minimum energy fitted, occurring on 25 stars (taken from their Table~6), and the elapsed time
of the observation was 660 ks; this converts to a flare rate of 51 flares yr$^{-1}$ star$^{-1}$ above a
minimum flare energy of 10$^{34}$ erg.
Table~\ref{tbl:calc} lists a mass-loss rate due to CMEs associated with these flares of 2$\times$10$^{-11}$ M$_{\odot}$ yr$^{-1}$,
and Table~\ref{tbl:extrap} lists a mass-loss rate due to flares in a common energy range
between 5$\times$10$^{31}$ and 10$^{37}$ erg of nearly twice that value.
These transient mass loss rates are elevated above the current total rate of solar mass loss by factors of $\approx$1000.

\subsubsection{EK Dra, a young solar analog}
The nearby (d=31 pc) G0V star EK~Dra has an age estimate of roughly 70 MY \citep{guinan2003}, and is 
generally considered to be a young solar analog.  Because of its proximity it has been
the subject of much attention.  
We calculate its escape velocity using the stellar mass and radius measurements listed in \citet{guinan2003}.
\citet{audard1999} used observations of EK~Dra with the {\it Extreme Ultraviolet Explorer (EUVE)}
to explore the rate of coronal flares in EK~Dra.
The energies determined in that paper represent the total coronal radiated energy of the flares
from 0.01-10 keV, and so we use the $f$ value in Table~\ref{tbl:f} appropriate for that waveband.
\citet{audard1999} fit a power-law between flare energies of 3$\times$10$^{33}$ and 3$\times$10$^{34}$ erg,
with a flare rate of 5$\times$10$^{-5}$ flares s$^{-1}$ above the minimum energy given above, or
a rate of 1550 flares yr$^{-1}$.

The mass loss rate listed in Table~\ref{tbl:calc} for CMEs accompanying these flares is 9$\times$10$^{-12}$
M$_{\odot}$ yr$^{-1}$, $\approx$ 450 times the total solar mass loss rate. 
When considering
a common flare energy range for mass loss determination, the implied mass loss rate is higher,
$\sim$4$\times$10$^{-11}$ M$_{\odot}$ yr$^{-1}$, or a factor of 2000 higher than the present day solar
mass loss rate. These rates suggest that the Sun might have been able to sustain a much higher rate of mass loss
in the past. 

\subsubsection{Superflaring Sun-like stars}
Recently \citet{maehara2012} reported on the incidence of flares seen on solar-like stars in Kepler
data.  The bandpass of the Kepler mission is relatively broad (4000-9000 \AA), and we do not know what
the spectral energy distribution of these flares is.  We assume that they are similar to solar and stellar flares in that
the optical flares dominate the energetics, and that the emission in the Kepler bandpass 
originates almost entirely from the hot continuum, with the same values of blackbody temperature
as seen in solar and stellar flares. 
Table~\ref{tbl:f} lists the fraction $f$ of the bolometric 
radiated energy that would appear in this bandpass under those assumptions.
Further characterization of these stars is ongoing, and we do not have precise radii or masses, so we take the
solar value of $v_{\rm esc}$.
For evaluation of equation~\ref{eqn:mdot}
 we use the number of flares (183) and the number of stars (86) that correspond
to the energy range fitted (5$\times$10$^{34}$--2$\times$10$^{36}$ erg) in the frequency distribution
(H. Maehara, priv. comm.). The data span 223 days, giving a flare rate of 3.5 flares yr$^{-1}$ star$^{-1}$.
This returns a mass loss rate of 8$\times$10$^{-13}$ for CMEs accompanying the flares considered (Table~\ref{tbl:calc}),
and $\dot M$ of 10$^{-11}$ M$_{\odot}$ yr$^{-1}$ for CMEs associated with 
a higher flare energy range (Table~\ref{tbl:extrap}).

\subsection{Low Mass Stars}
Low mass stars provide a stark contrast with flare behavior from higher mass solar-like stars:
the factor of $\sim$ three difference in  stellar mass and radius is accompanied by a very different internal structure.
The nearly or completely convective internal structure on low mass stars has profound implications for 
how magnetic fields are generated in the interior, and consequent implications for how that magnetic flux
emerges and interacts with plasma above the stellar surface.  As discussed above, detailed studies of individual
flares on M dwarfs do show agreement with solar flares. 
The long timescales for activity decay on M dwarfs
\citep{west2008} means that if CMEs accompany flares they have the potential to act as a significant mass loss 
process over a large fraction of the star's life.

\subsubsection{ Young Low Mass Stars in Orion}
\citet{caramazza2007} examined the flare frequency distributions of young low-mass stars in the Orion nebula
cluster from the {\it Chandra X-ray Observatory}.  
They studied stars with masses in the range 0.1--0.3M$_{\odot}$, and found no difference in the flare
rate for stars in this mass range compared to the higher mass solar-like stars in Orion studied by \citet{wolk2005}.
We assume that the spectral range over which the energies in \citet{caramazza2007}
are calculated is 0.5--8 keV \citep[the same as in][]{wolk2005} as this is not explicitly stated
in the paper. We also use the same fraction of radiated energy in this bandpass to 0.01-10 keV that we used
above for the young solar-mass stars in Orion, and the $f$ value in Table~\ref{tbl:f} for the soft X-ray
region to correct this energy to a bolometric value.
Using stellar parameters from \citet{getman2005} for this sample
we calculated a median escape velocity for the low mass stars in the
Caramazza et al. sample to be 214 km s$^{-1}$.
\citet{caramazza2007} studied 165 sources, and found 151 flares to which to fit a distribution.
Only flares containing 500 counts or more were contained in the distribution fit; this corresponds to an
energy of 10$^{34}$ erg. Their distribution extends to a maximum of $\sim$10$^{4}$ counts, giving a maximum energy
of 2$\times$10$^{35}$ erg. Converting these to the energies expected in the 0.01-10 keV range 
as in \S4.1.2 gives 2$\times$10$^{34}$ --
3$\times$10$^{35}$ erg. 
From their Figure~5, the normalization of their cumulative
distribution function at a value of 500 counts (their minimum flare energy to which to fit a power-law) is 0.2, for
0.2$\times$151 flares to consider. Their exposure time was the same as for the study of \citet{wolk2005} of 660ks.
Combining these, we arrive at a flare rate of 9 flares yr$^{-1}$ star$^{-1}$.
Table~\ref{tbl:calc} lists the calculated mass loss accompanying these flares as 3$\times$10$^{-12}$ M$_{\odot}$ yr$^{-1}$
using equation~\ref{eqn:mdot}.
Table~\ref{tbl:extrap} extrapolates to a mass loss between flares spanning 5$\times$10$^{31}$ and 10$^{37}$
erg of
2$\times$10$^{-11}$ M$_{\odot}$ yr$^{-1}$.

\subsubsection{Nearby Active M Dwarfs AD Leo and EV Lac}
The active M dwarfs AD~Leo and EV~Lac have exhibited extreme flaring activity in the past
\citep{hawleypettersen1991,osten2010}.  They are both nearby (d$\approx$5pc) and considered to be relatively
young \citep[age$\lesssim$300 MY;][]{shkolnik2009} with spectral types of M3 and M4 for AD~Leo and EV~Lac, respectively.
Because they are so active they have been the subject of many flare studies, and published flare frequency
distributions exist for both of these objects for both coronal and optical flares \citep{lme1976,audard2000}.
This offers an 
opportunity for a consistency check on the approach taken in the current paper.
We note that it was AD~Leo itself from which the $f$
values were calculated above in section 2, and so these numbers should be the most robust for 
consideration of its flares.  
We calculated escape velocities for the two stars using stellar data from \citet{reiners2009}.

\citet{audard2000} determined the flare frequency distribution for these two stars using EUVE observations,
and determined coronal flare energy values in the wavelength range 0.01-10 keV.
We used the value of $f$ listed in Table~\ref{tbl:f} for this bandpass.
They determined a flare rate of 7 flares day$^{-1}$ above a minimum energy 
of 6$\times$10$^{31}$ erg for AD~Leo, and
4 flares day$^{-1}$ above a minimum energy of 8$\times$10$^{31}$ erg for EV~Lac. These numbers lead to a total number of flares
of 2557 and 1461 flares yr$^{-1}$, 
and $\alpha$ values of
2.02$\pm$0.28, 1.76$\pm$0.33,
respectively, for AD~Leo and EV~Lac.
Table~\ref{tbl:calc} lists the values of mass-loss calculated for these
objects using equation~\ref{eqn:mdot}: 5$\times$10$^{-13}$ M$_{\odot}$ yr$^{-1}$  for AD~Leo and 4$\times$10$^{-13}$ M$_{\odot}$ yr$^{-1}$
for EV~Lac, and 3$\times$10$^{-13}$ M$_{\odot}$ yr$^{-1}$  for AD~Leo and 2$\times$10$^{-13}$ M$_{\odot}$ yr$^{-1}$
for EV~Lac using equation~\ref{eqn:mdot2}. There is good agreement between both M dwarfs as well as both techniques.

For these two M dwarfs, we also made use of the flare compilation of \citet{lme1976}, who 
determined flare frequency distributions for flares observed using U band photometry.  
We used the value of $f$ for the U band determined above and listed in Table~\ref{tbl:f}
to estimate the fraction of the total bolometric radiated flare energy appearing in the $U$ filter bandpass.
We estimated the minimum and maximum U band flare energies fitted in \citet{lme1976}
from the figures for each M dwarf (figures 13 and 14 from that paper); these are listed in Table~\ref{tbl:calc}.
We estimate the total number of flares per year occurring on each star in the energy range
applicable to their fitted flare frequency distributions (their equation 18) 
by using the $\alpha$ and $\beta$ parameters from their 
Table~3;
for AD~Leo this is 3790 flares yr$^{-1}$ above an E$_{\rm U,min}$ of 2$\times$10$^{30}$ erg
and for EV~Lac it is 1790 flares yr$^{-1}$ above an E$_{\rm U,min}$ of 1$\times$10$^{31}$ erg.
Converting between the differential \citep[$\alpha$, used here and in][]{audard2000}
and cumulative \citep[$\beta$, used in ][]{lme1976} forms of the flare frequency distribution,
and using the sign conventions appropriate to each, gives a transformation of
$\alpha=1-\beta$; the differential index is listed
in Table~\ref{tbl:calc}.
There is good agreement between the equipartition method and empirical method of estimating
flare-associated transient mass loss.

The mass loss rates computed from flare frequency distributions for optical and coronal flares listed
in Table~\ref{tbl:calc} for each of the two M dwarfs
differ from each other by about an order of magnitude (5$\times$10$^{-13}$ M$_{\odot}$ yr$^{-1}$,
6$\times$10$^{-14}$ M$_{\odot}$ yr$^{-1}$ for AD~Leo and 4$\times$10$^{-13}$ M$_{\odot}$ yr$^{-1}$,
9$\times$10$^{-14}$ M$_{\odot}$ yr$^{-1}$ for EV Lac).  However, the range of flare energies considered
in the coronal flare frequency distribution and the optical flare frequency distributions are quite different.
When converting to bolometric values, the coronal flare frequency distribution for AD~Leo covers the
range 2$\times$10$^{32}-$7$\times$10$^{33}$ erg while the U band flare frequency distribution covers the
range 2$\times$10$^{31}-$3$\times$10$^{32}$ erg. For EV Lac the bolometric flare energy ranges 
are 3$\times$10$^{32}-$7$\times$10$^{33}$ erg (coronal flares) and 9$\times$10$^{31}-$4$\times$10$^{32}$ erg (U band
flares). In both cases, the coronal flares are more energetic than the optical flares, and each covers only a small
range of energy (factor of 20-30 for the case of coronal flares, and factor of 4-15 for optical flares), but
put together and comparing $\alpha$ values,
they indicate a continuation of a similar trend over a larger energy region. 
Combining the two flare frequency distributions into a common energy range allows for a better comparison of the
results from the calculations done in this paper.
This is done in Table~\ref{tbl:extrap} where the agreement between the coronal and optical flare
estimates is much closer, $\sim$10$^{-12}$ M$_{\odot}$ yr$^{-1}$ for AD~Leo for both coronal and optical
flare frequency distributions, and 7$\times$10$^{-12}$ M$_{\odot}$ yr$^{-1}$,
10$^{-11}$ M$_{\odot}$ yr$^{-1}$ for EV~Lac from coronal and optical flare frequency distributions, respectively.

A different technique attempts to place boundaries on cool stellar mass loss, using the detection of astrospheric
absorption in the blue-ward wing of Lyman $\alpha$ emission as seen at high spectral resolution
and using this as an indirect means of detecting cool stellar mass loss
\citep{wood2005}. The astrospheric signature is interpreted using heliospheric models, part of which implies
a
mass loss rate for a solar-like stellar wind. 
The flare star EV~Lac is one of the targets with an astrospheric detection \citep{wood2005}.
Assuming spherical symmetry for a stellar wind and the applicability  of the heliospheric model 
applied to the astrospheric detection for 
this active M dwarf suggests a mass loss rate of 
$\dot M=2\times$10$^{-14}$ M$_{\odot}$ yr$^{-1}$.  
The extrapolated rate of transient mass loss for EV~Lac, using both the optical and coronal flare frequency
distributions, is about two orders of magnitude higher, according to Table~\ref{tbl:extrap}.
Both mass loss methods rely on extrapolations from the solar case, and their inconsistency likely reveals
where breakdowns occur between the two.  One immediate discrepancy is apparent in
examining the assumption of spatial distribution: the astrospheric models assume a spherically symmetric stellar wind,
whereas CMEs originate from a particular active region. 
As the CME expands outward, the motion approximates that of a segment of a spherical shell.
Thus, in the limit of a high rate of flares/CMES encompassing the whole surface, this has the
same effect as a steady stellar wind.


\subsubsection{Inactive M dwarfs}
\citet{hilton2011} observed flares in the U band on different types of M dwarfs, tabulating the flare frequency
distributions on early-, mid-, and late-M dwarf spectral types, 
for stars classified as magnetically active or inactive. A cutoff in characteristic H$\alpha$ emission level
was used to discriminate the two activity regimes.
Because there are not many studies of flare frequency distributions towards magnetically inactive stars, we use the tabulated
cumulative flare frequency distributions for 
inactive early- (M0-M2.5) and mid- (M3-M5) M dwarfs
to estimate the impact of flare-associated transient mass loss for these objects. 
It is important to point out that while these are inactive stars, they are still producing
flares with similar energies to those of active stars --- the main difference is the marked 
decrease in the flare rate.
In comparison with the very active M dwarfs described above, these stars provide a
view of the range of transient mass loss that might be expected from stars with lower
activity levels.

The flare frequency distribution is constructed in the same way as for the \citet{lme1976}
sample, and we use the same treatment as described above for the flare stars AD~Leo and EV~Lac. 
Flares from multiple stars were used to construct the distributions, and we corrected for this effect
in determining our $\dot N_{\rm tot}$. Using the parameters tabulated in \citet{hilton2011}'s Table 4.3, we
determine $\dot N_{\rm tot}$ of 26 flares yr$^{-1}$ star$^{-1}$ above a U band flare energy of 8$\times$10$^{29}$ erg for the inactive early M dwarfs
and 50 flares yr$^{-1}$ star$^{-1}$ above a U band flare energy of 4$\times$10$^{29}$ for the inactive mid-M dwarfs.  Table~\ref{tbl:calc}
lists the inferred mass-loss rate if coronal mass ejections accompany these flares. Due to the lower
flare rate the mass-loss rate is correspondingly lower, being 10$^{-16}$ $\dot M_{\odot}$ yr$^{-1}$ and
4$\times$10$^{-15}$ $\dot M_{\odot}$ yr$^{-1}$ for inactive early- and mid-M dwarfs, respectively, using
equation~\ref{eqn:mdot}, and 3$\times$10$^{-16}$ $\dot M_{\odot}$ yr$^{-1}$ for inactive early-M dwarfs,
2$\times$10$^{-15}$ $\dot M_{\odot}$ yr$^{-1}$ for inactive mid-M dwarfs using equation~\ref{eqn:mdot2}.
These levels are similar to those found for solar CME-related mass loss.

The $\alpha$ value for the inactive mid-M dwarfs is much lower than for flare frequency distributions
from the other types of stars considered: 1.2 compared to an average of 1.94$\pm$0.06
for the other distributions listed in Table~\ref{tbl:calc}.
The CME mass loss rate extrapolated to a common energy range tabulated in Table~\ref{tbl:extrap},
gives very different results.
This is due in large part to the different $\alpha$ values for the flare
frequency distributions.  
An $\alpha$ of 1.2 as for the inactive mid-M dwarfs
weighs larger flares more heavily, so extrapolating to an energy range
that encompasses much more energetic flares produces a higher $\dot M$. Since these stars were chosen specifically
to have a lower rate of overall magnetic activity as well as flare activity, it is not clear if
the upper flare energy is applicable. Thus the total mass loss rate calculated likely has a higher
level of uncertainty associated with it.

\section{Discussion}

\subsection{Investigation of Assumptions}
This analysis is predicated on the assumption that the large flares that are observed to occur on
stars are accompanied by large ejections of mass, and that we can use the
observed flare frequency distributions to infer the cumulative mass lost from these energetic events.
Flares manifest themselves in more than one region of the electromagnetic spectrum, stemming as they do
from structures that span a large vertical extent of the stellar atmosphere, from photosphere to corona. 
We have constructed a physically motivated way to connect flares and CMEs, which can be applied to flares
observed in different parts of the electromagnetic spectrum, by taking into account the fraction of the
total bolometric energy in that band.
The calculated mass loss rates in Table~\ref{tbl:calc} for magnetically active stars are large, while in comparison
the cumulative mass lost in solar CMEs is a small fraction of the total solar mass loss, as
expected from observations.  

There are several factors in the method that could affect the final calculation: 
the assumption for the flare
that $E_{\rm bol}=E_{\rm opt}+E_{\rm cor}$; the universality of energy partition from flare to flare; 
the assumption of equipartition between flare bolometric energy and CME kinetic energy ($\epsilon$=1);
the error on $\alpha$, the index of the flare frequency distribution; how studies define a flare, which affects $\dot N_{\rm tot}$; the assumption of $v=v_{\rm esc}$.
We discuss those here in turn.

\subsubsection{$E_{\rm bol}=E_{\rm opt}+E_{\rm cor}$}
Our assumption in tabulating the total bolometric flare energy is that the line and continuum emission in the
optical/UV, together with the coronal radiation, dominate the flare radiated energy budget.
The contribution to the radiative flare energy from radio and hard X-ray nonthermal radiation
will be negligible, due to the radiative inefficiency of nonthermal gyrosynchrotron and nonthermal
bremsstrahlung compared to collisional processes \citep{aschwanden2002}. 
It is possible that there are 
other components of flare radiation in relatively unexplored wavelength regions, but they likely do not dominate
the radiated flare energy budget.
\citet{euvflare} discussed a flare observed in the extreme ultraviolet with
a large release of energy in the 60-200 \AA\ bandpass, which also displayed enhanced continuum
emission in the 320--650 \AA\ range. The long wavelength continuum component contributed only at most 10\% of the estimated
total coronal radiated energy. 
\citet{flareatlas} described  results of a spectral atlas of M dwarf flares which showed the commonality of a 
continuum component appearing at wavelengths longer than about 4900 \AA. While this component becomes energetically more
important during the gradual phase of the flare, the hot blackbody at shorter wavelengths dominates the optical energy.

Because of the flow of energy through different parts of the atmosphere during a flare, it is 
worth considering whether energy is being double-counted by summing up different
radiated energy components into a bolometric radiated energy.
Radiative hydrodynamic modelling suggests that the black-body which figures prominently in the
optical continuum emission
arises because of nonthermal electrons hitting the lower atmosphere,
and theories of chromospheric evaporation \citep{fisher1985} 
explain the time correspondence between signatures which trace accelerated
particles (like the impulsive continuum emission in the optical, but also radio and nonthermal hard X-ray emission)
and signatures of coronal plasma heating
as the transfer of energy from the accelerated particles to heating plasma to coronal temperatures, the Neupert effect
\citep{neupert1968}.
Backwarming of coronal X-rays shining on the chromosphere/photosphere is likely to be negligible
\citep{allred2006}.
Additionally, we are not considering the kinetic energy of the accelerated particles in this
calculation, although other studies of solar and stellar flares have suggested that it could dominate
over the bolometric flare energies \citep{emslie2012,smith2005}.

\subsubsection{Universality of $f$}
By accounting for the fraction of the total flare radiated energy appearing in a given bandpass,
flare measurements made in different spectral regions are compared against each other and the implication
for flare-associated mass loss can be expanded.  There are relatively few stellar flares which have been observed using a
multiwavelength approach. 
\citet{hawleypettersen1991} demonstrated that the optical/UV flare
energy budgets of flares on M dwarfs appear to be similar to each other, so we are justified in
using the same fractions for different flaring M dwarfs.
The comparison of the flare energy partition values between these active stars and the Sun shows generally
good agreement, which we take as further justification for using the $f$ values derived for flares on different types
of stars. Further multiwavelength flares observations will be needed to investigate this assumption.

\subsubsection{Assumption of $\epsilon$=1 }
As discussed in \S 3, there is a reason to suggest a connection between the energetics of mass loss and
radiation. Solar eruptive events appear to bear this out.  \citet{emslie2012} described empirical
attempts to characterize the energy budget for 38 solar eruptive events, and find that E$_{\rm KE,CME}/$E$_{\rm flare,bol}$
$\sim$3. \citet{drake2013} also find a nearly linear relationship between the CME kinetic energy
and X-ray flare energy for a sample of CME-flare events. They restrict their analysis to the GOES energy band,
but find that E$_{\rm KE,CME}/$E$_{\rm GOES}$ $\sim$200 (their Figure~1), 
which reduces to E$_{\rm KE,CME}/$E$_{\rm bol}$ $\sim$2
after accounting for the 1\% fraction of bolometric radiated flare energy appearing in the GOES band.
\citet{emslie2005} noted that considering the uncertainties in the energy estimations, it was 
possible that the energies in the two components are roughly equal.
Taking a value for $\epsilon$ that is 0.5 or 0.3, as the solar studies of \citet{drake2013} and \citet{emslie2012}, respectively,
suggest, would lead to mass loss rates that are 2--3 times higher than what is calculated using Equation~\ref{eqn:mdot},
~\ref{eqn:mdot2} or their extrapolated values.

The association between flares and CMEs must break down at some low energy, where there isn't enough  magnetic free energy 
available for the system to erupt. This has been parameterized as the minimum energy at which the solar
flare/CME association departs from unity, but changing this number downward doesn't affect the numbers that much.
In reality, this limit will depend on other factors which will vary from flaring region to region.
The fact that the observed solar flare/CME relationship
departs from unity for smaller events demonstrates that there are physical limits on 
whether reconnection produces an eruption which can be observed as a CME.  
The approach put forward here concentrates on the impact
of the largest events.  

\subsubsection{Error on $\alpha$}

The uncertainty on the determination of $\alpha$ leads to a spread in the derived transient mass loss rate, but
it is a small factor.  For most of the published flare frequency distributions discussed earlier, varying the index
$\alpha$ of the distribution produced only a modest change in the derived mass loss rates. These range from as little
as a few percent difference to 60\% difference.  This effect is much smaller than the other factors affecting the method.
We do point out the singularity in equation~\ref{eqn:mdot} for a value of $\alpha=2$, and the singularity in equation
~\ref{eqn:mdot2} for a value of $\alpha=1+\gamma$ (or $\alpha \approx$1.6), both of which span the range of determined $\alpha$ values for stellar flares.
These equations can be recast to avoid this singularity in the appropriate limit ($\alpha \rightarrow 2$ 
or $\alpha \rightarrow 1+\gamma$) for large values of R. But this is not necessary because many of the 
studies cited here have only a modest
value of R (i.e. the U band compilation of flare frequency statistics for EV~Lac has a flare energy range R of only 5).  

\subsubsection{Definition of a flare} 

The various studies considered here used different methods to define a flare, and this will impact the
results primarily through the rate $\dot N_{\rm tot}$. The systematic effect these methods have on detecting the faint end of
the flare distribution will also sway the determination of $\alpha$, but inspection of Table~\ref{tbl:calc} shows a
range of $\alpha$ values generally in agreement with other studies quantifying the flare frequency distribution.  

Data obtained from optical light curves are binned, by the nature of the data collection method; the exposure times
set the noise level against which quiescent and flare emission is discerned, and thus the integrated energy/fluence
of flares.
\citet{hilton2011} used a series of consecutive flux measurements which are several $\sigma$ above a median quiescent flux,
for light curves with cadences of a few tens of seconds.
\citet{maehara2012} used a similar method to identify flares, looking at the distribution of brightness changes between pairs of
consecutive data points, and only looking at the top 1\% of this distribution as a way to filter out spurious candidates.
Since the Kepler light curves used in the \citet{maehara2012} study 
had integration times of nearly half an hour, this limits the flares to events lasting
longer than 1 hour.
\citet{lme1976} used a manual ``by-eye'' inspection of light curves to identify flares.

X-ray astronomical data consists of event lists, which are binned adapatively to search for flaring variations.
\citet{wolk2005} and \citet{caramazza2007} used maximum likelihood blocks to identify periods of elevated flux (without bias by 
time binning) then required that there be both a rapid rise to the elevated flux level, and that the elevated flux level
exceed the quiescent or characteristic level by at least 120\%.
\citet{audard2000} used an adaptive binning of light curves to determine the statistical distribution of counts in
quiescence; the beginning and ending of flares were identified as times corresponding to flux levels separated from the
maximum by 2 $\sigma$. They attempt to correct for overlapping flares.

Solar GOES light curves were binned in 1 minute cadence for the flare studies. 
The study of \citet{veronig2002}
did not correct for background emission in estimating flare energies.
They also note that their flare definition used four consecutive
points with increasing flux as part of the determination of the flare rise.  


All of these methods are sensitive to large flares, with the least sensitive only requiring a flare duration 
of an hour for the flare to be detected.

\subsubsection{Assumption of $v=v_{\rm esc}$}
In \S3.1, it was necessary to assume that the speed of the CME is constant and equal to 
the escape velocity of the star, in order to integrate the cumulative effect of the CMEs
through the observed flare frequency distribution. 
Solar CME speeds span a range from about 100 km s$^{-1}$ to 3000 km s$^{-1}$, 
with average flare-associated CME speed of 498 km s$^{-1}$
\citep{aarnio2011}. 
CME speeds on the Sun begin at velocities of $\sim$100 km s$^{-1}$
and range up to a few thousand km s$^{-1}$. 
Solar CMEs with speeds less than the solar wind speed experience acceleration up to the solar
wind speed, and fast speed CMEs decelerate to the solar wind speed, which is approximately the escape speed
\citep{cmeref}.
For any particular CME we could be off in estimating the
mass, by a factor ranging from a factor of about 25 lower to 40 higher using the solar CME speed range. 
But considering CMEs in aggregate,
the cumulative effect of mass loss should have a value which is closer to the ensemble average.
If stellar CMEs have larger average speeds, then this biases the total mass lost to a lower value.

Since the empirically derived stellar CME masses using equation~\ref{eqn:mdot2} and tabulated
in columns 9 and 10 of Table~\ref{tbl:calc}
have a functional dependence on 
flare energy which is less than linear (power-law exponent $\sim$0.6), the weight of large flares is less
using these methods than for the physically motivated model developed in this paper, where the dependence
of inferred CME mass on flare energy is linear. This is seen
in the $\dot M_{\rm Aarnio}$ and $\dot M_{\rm Drake}$ values in Table~\ref{tbl:calc} being
smaller than the $\dot M$ values derived above for flare energy distributions which 
go out to large flare energies.
Apart from very active stars such as these examples (the young suns in Orion, superflaring sun-like stars, and young low mass stars)
which have
E$_{\rm rad,max} >$10$^{35}$ erg, the agreement between the two approaches is reasonably good.
Given the many uncertainties involved in making an association between flares and CMEs, we consider the
general agreement between this method and the empirical method in \citet{aarnio2012} and \citet{drake2013}
to be a good sign. 


\subsection{Astrophysical consequences of high mass loss rates}

The mass loss rates suggested by this analysis are large for magnetically active flaring stars, and
have a number of astrophysical consequences. 
\citet{drake2013} considered the effect of CMEs associated with stellar flares in a flare-heated corona, and
determined a high level of kinetic energy required, in excess of the radiative heating input.  
A nonstandard solar model with a high mass-loss rate might be one solution to the faint young
Sun paradox \citep{fys2007}.
The results from Table~\ref{tbl:calc} for the young Suns in the
Orion Nebula Cluster and the young solar analog EK~Dra suggest that flare-associated transient mass loss
could have been substantial for at least the first $\sim$100 MY of the Sun's history,
although the mass loss rate values derived here are lower than what
is necessary to change the young Sun's mass significantly.
Thus the total mass lost during the active phases of these stars is not significant enough to 
alter their evolution.

An enhanced mass-loss rate also has implications for the stellar environment.
Flares which connect to the planet-forming disk could be important in regulating and/or influencing planet formation,
and 
coronal mass ejections might also impact the disk in young stars, resulting in disk seeding with 
processed stellar material. This could help explain aspects of radionuclide production in protoplanetary disks \citep{feigelson2002}.
The transient mass loss acts as an enhanced stellar wind, and might influence the planetary dynamo \citep{heyner2012}.
CMEs might also play a role in removing material from debris disks and producing the recently observed dramatic
IR variability, as explained by 
\citep{osten2013}.
\citet{colemanworden1976} calculated that for a stellar mass
loss of about $6\times10^{-12}$ $M_{\odot}$ yr$^{-1}$ from M dwarfs, a significant amount of ISM material would be supplied 
by M dwarfs.  
However, even with this high rate of mass-loss, and the abundance of M dwarfs in the galaxy, the  
cumulative mass lost is much less than that contributed by the rarer low-mass red giant stars \citep{deBoer2004}.
At young ages for planets around solar-like stars in the habitable zone, 
the enhanced transient mass loss might lead to an enhanced compression of the
planetary magnetosphere and exposure of the exoplanet
atmosphere to ionizing radiation.
Around low-mass stars, this potential effect is magnified due to the proximity of the habitable zone to 
the star
\citep{khodachenko2007}.

The reality of such high inferred mass loss rates has been questioned before. \citet{drake2013} inferred
mass-loss rates as high as 10$^{-10}$ M$_{\odot}$ yr$^{-1}$ for active stars, requiring up to 10\% of the
stellar bolometric luminosity. 
\citet{limwhite1996} used upper limits on millimeter-wavelength radio fluxes to argue
that winds from active stars could be no more than 1-2 orders of magnitude higher that the present-day
solar mass-loss rate, in order that the observed centimeter-wavelength (and longer) radio flares
not be obscured by the optically thick wind. 
Coronal magnetic topologies on both pre-main sequence stars and more evolved active main
sequence stars show evidence of a complex geometry compared to the Sun \citep{donati2009},
and a reduction in the fraction of open versus closed field lines might have a concomitant decrease in the amount
of mass loss from a coronal wind. 
On the other hand, \citet{cohen2009} demonstrated that polar spots on active stars
could enhance the mass loss rate from a steady wind, albeit in an asymmetric manner.

There is a complex magnetic topology involved in active regions on stars, and the specific configuration
may provide an explanation as to whether a particular reconnection event could result in an ejection. 
Strong overlying fields might participate in the reconnection, or alternatively provide a barrier to eruption.
On stars with 
stronger surface fields and denser coronal plasma than the Sun this could be difficult to achieve.
On this point we ought to be able to gain insight from the Sun: a case study of AR 12192, one of the largest
active regions seen in decades, which produced several energetic flares in the October-November 2014
time frame without accompanying CMEs, can reveal the detailed processes
for noneruptive energetic flares.
Preliminary work on failed solar eruptive events by \citet{chen2013} and \cite{sun2015} does suggest that overlying arcades may prevent the eruption of
a filament.
In the end, only direct observations of CMEs in other stars (or more stringent constraints) can address whether and how the above topics
are influenced by transient mass loss.
In this regard, observational signatures uniquely associated with the CME which are detectable in nearby
magnetically active stars provide a path forward.  
The radio signature of coronal mass ejections on the Sun is the so-called type II burst, which has a distinct signature
of a slowly drifting radio burst. 
The type II burst arises from plasma emission in the tenuous outer atmosphere of the star,
produced as the result of MHD shocks from the passage of the fast-moving CME through the atmosphere.
From a study of the dynamic spectrum of such a slowly drifting burst, the velocity of the
shock can be extracted (providing an upper limit on the CME speed) 
and with an assumption of equipartition, the mass of the CME
derived.
A direct measurement of a slowly drifting stellar metric radio burst would be another step in
validating the solar/stellar connection with respect to the flare process, and would establish a
relatively unambiguous observational technique to study the influence of transient mass loss in
active cool stars.

\section{Summary and Conclusions}


We established an energy partition for stellar
flares, which allows for an intercomparison of flares observed in different regions of the
electromagnetic spectrum. Using this energy partition, and applying an equipartition
between CME kinetic energy and flare bolometric energy, we investigated the implication
of CMEs associated with these flares for a variety of different kinds of stars having
measured flare frequency distributions. 
We relate the cumulative transient mass loss to the total flare rate, a point which has not been
emphasized in previous studies.
While empirical or scaling relations have been used in the past to connect flares and mass ejections
on stars, the recent details of energy budgets in solar eruptive events allows
for a physical explanation for connecting the two processes.
We also used empirical correlations established between solar flare X-ray energy and solar
CME mass to derive transient mass loss rates.
The implied rate of transient mass loss, for both methods, is generally
large.
Our general method also allows for an intercomparison of flare frequency distributions for the 
same object taken in different wavebands; the comparison done for two active M dwarfs shows good agreement.
There are several important consequences to an increase in the transient mass loss
from magnetically active stars, and   
observational constraints on the presence of stellar CMEs and their relation to
stellar flares would be invaluable to evaluating their impact.
Just as valuable would be a refutation of the solar-stellar connection in this
aspect of magnetic reconnection.

A comparison of implications for flare frequency distributions of the same star at two different
wavelength regions reveals consistency, which is reassuring. This means that the distributions are
sampling the same intrinsic flare distribution. The flare frequency distribution is commonly
used in coronal studies to examine the coronal heating hypothesis (suggested for cases where the
differential $\alpha$ exceeds 2), but their consistency between optical and coronal means that such
studies are able to be carried out with optical telescopes. 
The results of the energy partition calculations in Table~\ref{tbl:f} are used to relate the
amount of radiated flare energy in one bandpass to the total bolometric amount.
In particular, the long time baselines allowed by
space missions like Kepler, K2, and TESS can inform coronal studies, especially in situations of infrequent
but large flare energy releases.
Such missions will increase the yield of flares studied on different classes of stars,
at the same time expanding the stellar types for which transient mass loss can be investigated.


\acknowledgements
RAO acknowledges the useful discussions which took place at the International Space Science Institute
workshop on Energy Partition in Solar and Stellar Flares for helping to frame elements of the study.
SJW was supported by NASA contract NAS8-03060 (Chandra).


\end{document}